\documentclass[12pt]{article} 
\usepackage{amssymb}
\usepackage{graphicx}
\begin{document}  
\begin{center}
{\large \bf The interaction region of high energy protons}

\vspace{0.5cm}                   

{\bf Igor M. Dremin$^{1,2}$, Sebastian N. White$^3$}

\vspace{0.5cm}                       

        $^1$Lebedev Physics Institute, Moscow 119991, Russia\\
\medskip

    $^2$National Research Nuclear University "MEPhI", Moscow 115409, Russia     

\medskip 

   $^3$CERN, CH-1211 Geneva 23, Switzerland
 
\end{center}

\begin{abstract}
The spatial view of the interaction region of colliding high energy protons (in terms of impact parameter)
is considered. It is shown that the region of inelastic collisions has a
very peculiar shape. It saturates for central collisions at an energy of 7 TeV.
We speculate on the further evolution with energy, which is contrasted to the  
"black disk" picture.
\end{abstract}

\section{Introduction}

The search ever deeper into the interior of matter successfully started by Rutherford's
discovery of atomic structure is going on now at much lower scales (below
10$^{-13}$ cm) at high energy accelerators. The interaction region of colliding
protons can be quantitatively explored with the help of the unitarity condition if 
experimental data on their elastic scattering are used. With only these
two ingredients at hand we are able to show that the energy evolution of the
inelastic interaction region demonstrates quite surprising features.                                            

\section{The unitarity condition}
From the theoretical side, the most reliable information comes from the 
unitarity condition. The unitarity of the 
$S$-matrix $SS^+$=1 relates the amplitude of elastic
scattering $f(s,t)$ to the amplitudes of inelastic processes $M_n$. 
In the $s$-channel they are subject to the integral relation (for more
details see, e.g., \cite{PDG, anddre1, ufnel}) which can be written 
symbolically as
\begin{equation}
{\rm Im}f(s,t)= I_2(s,t)+g(s,t)=
\int d\Phi _2 ff^*+\sum _n\int d\Phi _n M_nM_n^*.
\label{unit}
\end{equation}
The variables $s$ and $t$ are the squared energy and transferred momentum of 
colliding protons in the center of mass system $s=4E^2=4(p^2+m^2)$, 
$-t=2p^2(1-\cos \theta)$ at the scattering angle $\theta $.  
The non-linear integral term represents the two-particle intermediate states of the 
incoming particles. The second term represents the shadowing contribution of 
inelastic processes to the imaginary part of the elastic scattering 
amplitude. Following \cite{hove} it is called the overlap function. This 
terminology is ascribed to it because the integral there defines the overlap 
within the corresponding phase space $d\Phi _n$ between the matrix element 
$M_n$ of the $n$-th inelastic channel and its conjugated counterpart with the 
collision axis of initial particles deflected by an angle $\theta $ in proton 
elastic scattering. It is positive at $\theta =0$ but can 
change sign at $\theta \neq 0$ due to the relative phases of inelastic matrix 
elements $M_n$'s.

At $t=0$ it leads to the optical theorem 
\begin{equation}
{\rm Im}f(s,0)=\sigma _{tot}/4\sqrt {\pi}
\label{opt}
\end{equation}
and to the general statement that the total cross section is the sum of 
cross sections of elastic and inelastic processes
\begin{equation}
\sigma _{tot}=\sigma _{el}+\sigma _{in},
\label{telin}
\end{equation}
i.e., that the total probability of all processes is equal to one.

\section{The geometry of the interaction region}

Here, we show that it is possible to study the space structure of the 
interaction region of colliding protons using information about their 
elastic scattering within the unitarity condition. The whole procedure is
simplified because in the space representation one gets an algebraic
relation between the elastic and inelastic contributions to the unitarity
condition in place of the more complicated non-linear integral term 
$I_2$ in Eq. (\ref{unit}).

To define the geometry of the collision we must express all characteristics
presented by the angle $\theta $ and the transferred momentum $t$
in terms of the transverse distance between the trajectories of the centers 
of the colliding protons - namely the impact parameter, $b$. This is easily carried out using the 
Fourier -- Bessel transform of the amplitude $f$ which retranslates the
momentum data to the corresponding transverse space features and is written as
\begin{equation}
i\Gamma (s,b)=\frac {1}{2\sqrt {\pi }}\int _0^{\infty}d\vert t\vert f(s,t)
J_0(b\sqrt {\vert t\vert }).
\label{gamm}
\end{equation}
The unitarity condition in the $b$-representation reads
\begin{equation}
G(s,b)=2{\rm Re}\Gamma (s,b)-\vert \Gamma (s,b)\vert ^2.
\label{unit1}
\end{equation}
The left-hand side (the overlap function in  the $b$-representation) describes the 
transverse impact-parameter profile of inelastic collisions of protons. It is 
just the Fourier -- Bessel transform of the overlap function $g$. It satisfies 
the inequalities $0\leq G(s,b)\leq 1$ and determines how absorptive the 
interaction region is, depending on the impact parameter (with $G=1$ for full 
absorption and $G=0$ for complete transparency). The profile of elastic 
processes is determined by the subtrahend in Eq. (\ref{unit1}). If $G(s,b)$ is
integrated over all impact parameters, it leads to the cross section for 
inelastic processes. The terms on the right-hand side would produce the total
cross section and the elastic cross section, correspondingly, as should be the case
according to Eq. (\ref{telin}). The overlap 
function is often discussed in relation with the opacity (or the eikonal phase) 
$\Omega (s,b)$ such that $G(s,b)=1-\exp (-\Omega (s,b))$. Thus, full
absorption corresponds to $\Omega =\infty $ and complete transparency to
$\Omega =0$. 
 
The most prominent feature of elastic scattering is the rapid decrease of the 
differential cross section with increasing transferred momentum,
$\vert t\vert $, in the diffraction peak. As a first approximation, at present 
energies, it can be described by the exponential shape with the slope $B(s)$:
\begin{equation}
\frac {d\sigma }{dt}=\frac {\sigma ^2_{tot}}{16\pi }\exp (-B(s)\vert t\vert ).
\label{expB}
\end{equation}
The diffraction cone contributes predominantly to the Fourier - Bessel transform of the
amplitude. Using the above formulae, one can write the dimensionless $\Gamma $ 
as
\begin{equation}
i\Gamma (s,b)=\frac {\sigma _t}{8\pi }\int _0^{\infty}d\vert t\vert 
\exp (-B\vert t\vert /2 )(i+\rho )J_0(b\sqrt {\vert t\vert }).
\label{gam2}
\end{equation}
Here, the diffraction cone approximation (\ref{expB}) is inserted. 
Herefrom, one calculates
\begin{equation}
{\rm Re}\Gamma (s,b)= {\zeta }{\exp (-\frac {b^2}{2B})},
\label{rega}
\end{equation}
where we introduce the dimensionless ratio of the cone slope (or the elastic
cross section) to the total cross section
\begin{equation}
\zeta=\frac {\sigma _{tot}}{4\pi B}= \frac {4\sigma _{el}}{(1+\rho ^2)\sigma _{tot}} \approx \frac {4\sigma _{el}}{\sigma _{tot}}.
\label{ze}
\end{equation}
The ratio $\sigma _{el}/\sigma _{tot}$ defines the survival probability of
initial protons. The approximation sign refers to the neglected factor
$1+\rho ^2$ where $\rho $ is the ratio of the real to imaginary part of the 
amplitude in the diffraction cone. In what follows we neglect $\rho $  
according to experimental data (with $\rho (7 \; TeV, 0)\approx 0.145$) and 
theoretical considerations which favor its decrease inside the diffraction cone.
Thus one gets
\begin{equation}
G(s,b)=  \zeta \exp (-\frac {b^2}{2B})[2-\zeta \exp (-\frac {b^2}{2B})].
\label{ge}
\end{equation}
The inelastic profile depends on two measured quantities - the diffraction cone width
$B(s)$ and its ratio to the total cross section, $\zeta $. It scales as a 
function of $b/\sqrt {2B}$.

For central collisions with $b=0$ one gets
\begin{equation}
G(s,b=0)= \zeta (2-\zeta).
\label{gZ}       
\end{equation}
This formula is very significant because it follows herefrom that the darkness 
at the very center is fully determined by only one parameter, $\zeta $, 
which is the ratio of experimentally measured quantities. It is given by
the ratio of the width of the 
diffraction cone $B$ (or $\sigma _{el}$) to the total cross section. The energy 
evolution of these quantities defines the evolution of the absorption value.
The interaction region becomes completely absorptive $G(s,0)=1$ in the center 
only at $\zeta =1$ and the absorption diminishes for other values of $\zeta $.
However for small variations of $\zeta =1\pm \epsilon $ the value of 
$G(s,0)=1-\epsilon ^2$ varies even less.

In the Table, we 
show the energy evolution of $\zeta $ and $G(s,0)$ for $pp$ and $p\bar p$ 
scattering as calculated from experimental data about the total cross section
and the diffraction cone slope at corresponding energies.
\medskip
\begin{table}
\medskip
Table.  $\;\;$ The energy behavior of $\zeta $ and $G(s,0)$.
\medskip

    \begin{tabular}{|l|l|l|l|l|l|l|l|l|l|l|l}
        \hline
$\sqrt s$, GeV&2.70&4.11&4.74&7.62&13.8&62.5&546&1800&7000\\ \hline
$\zeta $           &1.56&0.98&0.92&0.75&0.69&0.67&0.83&0.93&1.00-1.02 \\  
$G(s,0)$     &0.68&1.00&0.993&0.94&0.904&0.89&0.97&0.995&1.00 \\  \hline
   
\end{tabular}
\end{table}
Let us point out that starting from ISR energies the value of $\zeta $ increases
systematically and at LHC energies becomes equal to 1 within the accuracy of
measurements of $B$ and $\sigma _{tot}$.

The impact parameter distribution of $G(s,b)$ (\ref{ge}) has its maximum 
at $b_m^2=2B\ln \zeta$ with full absorption $G(b_m)=1$. 
Its position depends both on $B$ and $\zeta $.

Note, that, for $\zeta <1$ (which is the case, e.g., at ISR energies) one gets 
incomplete absorption $G(s,b)<1$ at any physical $b\geq 0$ with the largest 
value reached at $b=0$ because the maximum appears at non-physical values of 
$b$. The disk is semi-transparent. 

At $\zeta =1$, which is reached at 7 TeV, the maximum is positioned exactly 
at $b=0$, and full absorption occurs there, i.e. $G(s,0)=1$. The disk center 
becomes impenetrable 
(black). The strongly absorptive core of the inelastic interaction region grows 
in size as we see from expansion of Eq. (\ref{ge}) at small impact parameters:
\begin{equation}
G(s,b)= \zeta [2-\zeta -\frac {b^2}{B}(1-\zeta )-\frac {b^4}{4B^2}(2\zeta -1)].
\label{gb}
\end{equation}
The term proportional to $b^2$ vanishes at $\zeta =1$, and $G(b)$ 
develops a plateau which extends to quite large values of $b$ (about 0.5 fm). The 
plateau is very flat because the last term starts to play a role at 7 TeV 
(where $B\approx 20$ GeV$^{-2}$) only for larger values of $b$. 

At $\zeta >1$, the maximum shifts to positive physical impact parameters. A
dip is formed at $b$=0 leading to a concave shaped inelastic 
interaction region - approaching a toroidal shape. This dip becomes deeper at 
larger $\zeta $. The limiting value $\zeta =2$ leads to complete 
transparency at the center $b=0$.

\begin{figure}
\caption{ The evolution of the inelastic interaction region in terms 
of the survival probability. The values $\zeta =0.7$ and $1.0$ correspond to 
ISR and LHC energies and agree well with the result of detailed fitting to the 
elastic scattering data \cite{amal, dnec, mart}. 
A further increase of $\zeta $ leads to the toroid-like shape with a dip at $b=0$.
The values $\zeta =1.5$ are proposed in \cite{kfk, fms} and $\zeta =1.8$ in
\cite{roy} as corresponding to asymptotical regimes. The value $\zeta =2$
corresponds to the "black disk" regime ($\sigma _{el}=\sigma _{inel}=
0.5\sigma _{tot}$).}
\centerline{\includegraphics[width=\textwidth, height=9cm]{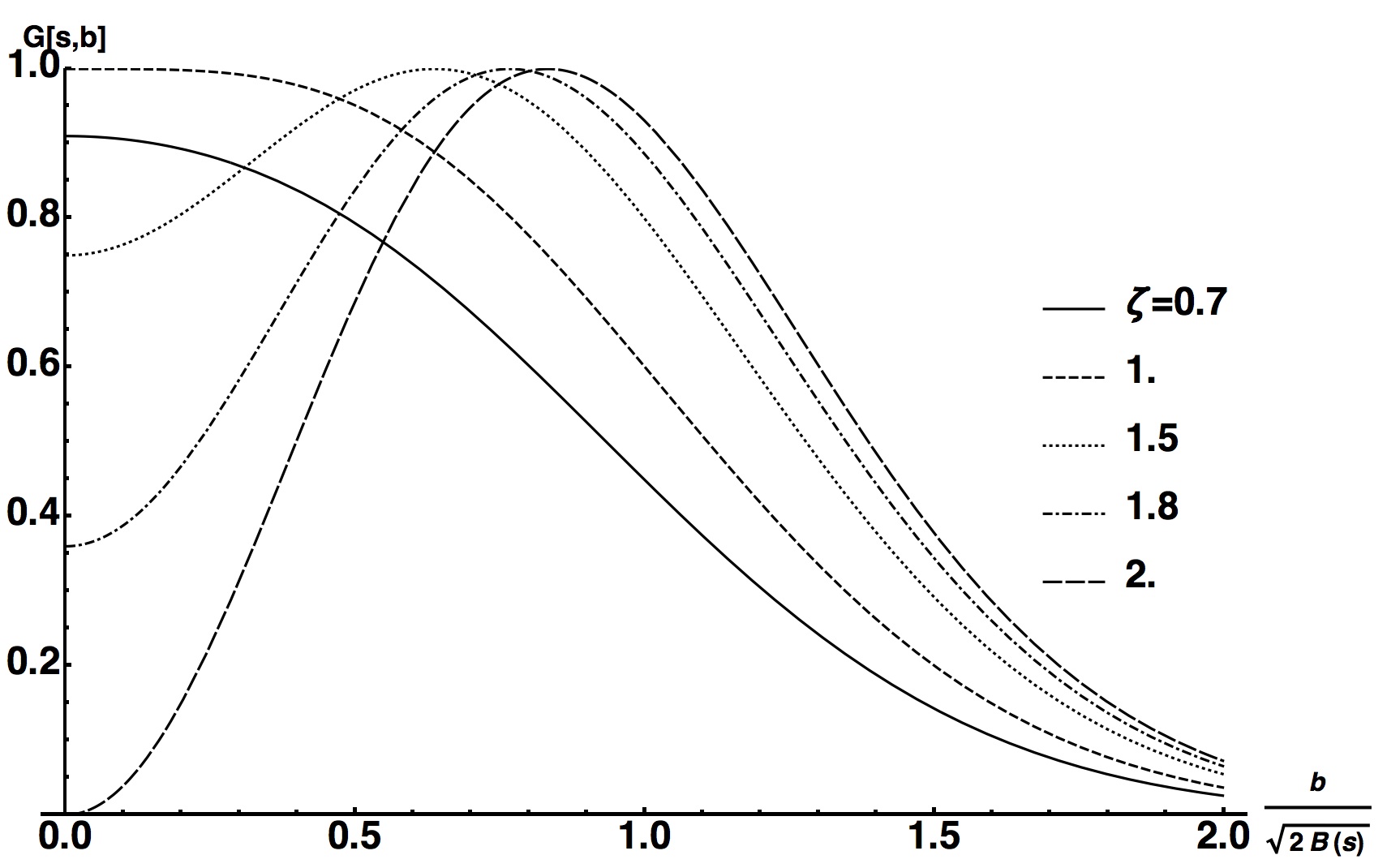}}

\end{figure}
All these cases are demonstrated in Fig. 1 where $G(s,b)$ is plotted as a 
function of the scaling variable $b/\sqrt {2B}$ for different values of the
parameter $\zeta $ according to Eq. (\ref{ge}). The line with $\zeta =0.7$
corresponds to ISR results and with $\zeta =1$ to LHC. Earlier it was shown 
that the results of analytical calculations according to (\ref{ge}) and the 
computation with experimental data directly inserted in the unitarity condition 
practically coincide (see Fig. 1 in \cite{ads}). 

What can we expect at higher energies?

The profiles shown in Fig. 1 are valid so long as we can assume that the
differential cross section of elastic scattering decreases exponentially 
with $\vert t\vert $ within the diffraction cone.
They can change if this traditional behavior
is no longer valid at higher energies. Slope variations of the order
of 1 per cent found at 8 TeV by TOTEM \cite{totem} are still not significant.

Only guesses can be obtained from the extrapolation of results at lower 
energies to new regimes, even though experience shows how indefinite and 
even erroneous such extrapolations can be.

First, one may assume that $\zeta $ will increase without crossing 1 but 
approaching it asymptotically. That would imply that its precise value at 7 TeV 
is still slightly lower than 1 within the present experimental errors\footnote{ The value of
$\sigma _{el}/\sigma _{tot}=0.257\pm0.005$ reported in \cite{totem1} would imply $\zeta =1.01$ with an
uncertainty of $\sim2\%$.} . Then the 
inelastic profile shown in Fig. 1 for $\zeta =1$ will be quite stable with a
slow approach to complete blackness in central collisions and a steady 
increase of its range. This situation seems most appealing to our 
theoretical intuition. 

However, given the experimentally observed increase of the share
of elastic scattering from ISR to LHC, it is tempting to consider another 
intriguing possibility- that there could be a further increase at still higher energy. Then the interaction region 
inevitably acquires a toroid-like shape with a dip at the very center ($b=0$). 
Some extrapolations of fits at lower energies are presented in \cite{kfk, fms}
and theoretical speculations are discussed in \cite{roy}.
The line with $\zeta =1.5$ describes the profile of the inelastic 
interaction region according to asymptotic 
expectations predicted in \cite{kfk, fms},  where successful fits to present
experimental data are reported. The new proposal of \cite{roy} is shown at
$\zeta =1.8$. The dip increases at larger $\zeta $ and reaches the very bottom
$G(0)=0$ for $\zeta =2$. Strangely enough this situation with $\sigma _{el}=
\sigma _{inel}=0.5\sigma _{tot}$ is usually referred to as the "black disk" 
limit \cite{blha}.

Protons become impenetrable when  $b=0$ for $\zeta =2$ and only undergo
elastic scattering. It is discussed in Ref. \cite{trt}. This condition
results in purely backward scattering (as in head-on collisions 
of billiard balls). 

In conclusion, we can state that, analyzing the unitarity condition,
we have found a special role for the ratio of elastic to total
cross sections being equal to 1/4 in 7 TeV pp-interactions
and described the consequences of its energy evolution.
This role could be attributed to an equal share of processes with
exchange and no-exchange of quantum numbers in particle collisions.
Then elastic processes constitute a half of the no-exchange share. Another 
half would be attributed to inelastic diffraction processes. That would 
lead to saturation of the Pumplin bound \cite{pump}
which states that their sum is less or equal to 0.5 of the total
cross section. However there is still no consensus among experiments
about the saturation of the bound at 7 TeV (see \cite{CMS, ALICE}).
Although there is currently a large latitude for the inelastic diffractive 
cross sections permitted by the accuracy of the experiments, as was pointed out
in \cite{lilu} the value presented in \cite{ALICE} corresponds to perfect 
agreement with the above picture( ie the above-metioned saturation).

In general, inelastic diffraction is determined by the dispersion of matrix
elements while only their averages enter into Eqs (10), (11). Some models have 
to be invoked in order to predict the dispersion. On a qualitative level it 
looks as though the absorptive structure of protons is extremely inhomogeneous
\cite{fimi}. That could explain the behavior of the inelastic profile
described above.

\medskip

{\bf Acknowledgments}

\medskip 
 
I.D. is grateful for support by the RFBR-grant 14-02-00099 and 
the RAS-CERN program.


\begin{thebibliography}{99}
\bibitem{PDG}
PDG group, China Phys. C  {\bf 38}, 090513 (2014).
\bibitem{anddre1}
I.V. Andreev, I.M. Dremin, ZhETF Pis'ma {\bf 6} (1967) 810
\bibitem{ufnel}
I.M. Dremin, Physics-Uspekhi {\bf 56} (2013) 3; {\bf 58} (2015) 61
\bibitem{hove}
L. Van Hove, Nuovo Cimento {\bf 28} (1963) 798
\bibitem{amal}
U. Amaldi, K.R. Schubert, Nucl. Phys. B {\bf 166} (1980) 301
\bibitem{dnec}
I.M. Dremin, V.A. Nechitailo, Nucl. Phys. A {\bf 916} (2013) 241.
\bibitem{mart}
A. Alkin, E. Martynov, O. Kovalenko, S.M. Troshin, Phys. Rev. D {\bf 89}
(2014) 091501(R)
\bibitem{kfk}
A.K. Kohara, E. Ferreira, T. Kodama, pp elastic scattering at LHC energies;
arXiv:1408.1599
\bibitem{fms}
D.A. Fagundes, M.J. Menon, P.V.R.G. Silva, Exploring central opacity
and asymptotic scenarios in elastic hadron scattering; arXiv:1509.04108
\bibitem{roy}
S.M. Roy, A two component picture for high energy scattering: unitarity, 
analitycity and LHC data; arXiv:1602.03627
\bibitem{ads}
M.Yu. Azarkin, I.M. Dremin, M. Strikman, 
Phys. Lett. B {\bf 735} (2014) 244; arXiv:1401.1973
\bibitem{totem}
TOTEM Collaboration, Nucl. Phys. B {\bf 899} (2015) 527
\bibitem{totem1}
TOTEM Collaboration, EPL {\bf 101} (2013) 21004
\bibitem{blha}
M.M. Block, F. Halzen, Phys. Rev. Lett. {\bf 107} (2011) 212002
\bibitem{trt}
S.M. Troshin, N.E. Tyurin, Phys. Lett. B {\bf 316} (1993) 175
\bibitem{pump}
J. Pumplin, Phys. Rev. D {\bf 8} (1973) 2899
\bibitem{CMS}
CMS Collaboration, Phys. Rev. D {\bf 92} (2015) 012003
\bibitem{ALICE}
ALICE Collaboration, EPJ C {\bf 73} (2013) 2456 
\bibitem{lilu}
P. Lipari, M. Lusignoli, EPJ C {\bf 73} (2013) 2630
\bibitem{fimi}
K. Fialkowski, H.I. Miettinen, Nucl. Phys. B {\bf 103} (1976) 247
\end{thebibliography}
\end{document}